A Field-Mill Proxy Climatology for the Lightning Launch Commit Criteria at Cape Canaveral Air Force Station and NASA Kennedy Space Center

Shane Gardner[1], Edward White[1,*], Brent Langhals[2], Todd McNamara[3], William Roeder[4], and Alfred E. Thal, Jr.[2]

\*       Corresponding author

1       Department of Mathematics and Statistics, Air Force Institute of Technology Wright-Patterson Air Force Base, Ohio

2       Department of Systems Engineering and Management, Air Force Institute of Technology, Wright-Patterson Air Force Base, Ohio

3       Architecture and Integration Staff Meteorology Division, Air Force Life Cycle Management Center, Wright Patterson Air Force Base, Ohio

4       U.S. Air Force 45th Weather Squadron, Cape Canaveral Air Force Station, Florida

# Abstract

The Lightning Launch Commit Criteria (LLCC) are a set of complex rules to avoid natural and rocket-triggered lightning strikes to in-flight space launch vehicles. The LLCC are the leading source of scrubs and delays to space launches from Cape Canaveral Air Force Station (CCAFS) and NASA Kennedy Space Center (KSC). An LLCC climatology would be useful for designing launch concept of operations, mission planning, long-range forecasting, training, and setting LLCC improvement priorities. Unfortunately, an LLCC climatology has not been available for CCAFS/KSC. Attempts have been made to develop such a climatology, but they have not been entirely successful. The main shortfall has been the lack of a long continuous record of LLCC evaluations. Even though CCAFS/KSC is the world's busiest spaceport, the record of LLCC evaluations is not detailed enough to create the climatology. As a potential solution, the research in this study developed a proxy climatology of LLCC violations by using the long continuous record of surface electric field mills at CCAFS/KSC.

A Field-Mill Proxy Climatology for the Lightning Launch Commit Criteria at Cape Canaveral Air Force Station and NASA Kennedy Space Center

1. Introduction and Background

The primary weather challenge at the Cape Canaveral Air Force Station and Kennedy Space Center (CCAFS/KSC) is lightning (Mazany et al. 2002). Lightning has many impacts on operations at CCAFS/KSC, including lightning warnings for personnel safety and resource protection, lightning forecasts for ground processing operations preparing for space launch, lightning forecasts for exposure forecasts during launch windows, and daily lightning reports to assess the risk of induced current damage in electronics of satellite payloads, rocket avionics, and test equipment. One of the largest lightning impacts is when weather conditions violate the Lightning Launch Commit Criteria (LLCC), the complex weather rules to avoid natural and rocket-triggered lightning strikes to in-flight space launch vehicles. Rocket-triggered lightning can occur when a space vehicle travels through an area of the atmosphere with an amplified electrical field. This, coupled with the vehicle's exhaust plume, can increase the electric field strength surrounding the vehicle (Mazur 2016). As a result, electric fields that are not strong enough to cause natural lightning become strong enough under the influence of the rocket for lightning to occur—this is referred to as rocket-triggered lightning.

The LLCC negatively impact space operations at CCAFS/KSC, but they are needed to assure safety. Approximately one-third of all space launch countdowns are delayed or scrubbed due to natural or triggered lightning threats or both (Hazen et al. 1995).  To ensure the operation of the flight termination system for space launch vehicles, the 45 Weather Squadron (WS) adheres to the LLCC for space launches.  The flight termination system is used to destroy the in-flight rocket if it goes outside of the limits for allowable flight path.  This is done to assure public safety.  If the in-flight space launch vehicle were to suffer a natural or rocket-triggered lightning strike, the flight termination system could be damaged, thereby stopping the ability to terminate the flight if needed.

This set of LLCC rules has evolved over the years as various elements of meteorological risks have been identified and mitigation steps adopted.  Apollo XII initiated the National Aeronautical Space Association's (NASA's) investigation into rocket-triggered lightning when this launch triggered two lightning strikes, causing a temporary loss of power and controls which almost brought an abrupt end to the mission. Fortunately, adequate backup systems allowed the flight to proceed without disaster; however, increased attention to triggered lightning soon followed (Kemppanen 2019). While the almost disaster for the three astronauts in Apollo XII initiated investigation into rocket-triggered lightning, the watershed moment occurred in 1987.  Rocket-triggered lightning caused the space launch vehicle to change its orientation, which led to the activation of the flight termination system and the resultant destruction of the Atlas Centaur (AC) 67 and its payload.

In 1988, the Lightning Review Committee (currently known as the Lightning Advisory Panel (LAP)), an independent review board of atmospheric electricity experts, formally developed the LLCC, a set of rules that prohibit a space launch if any one of twelve established criteria is violated.  One of these rules, the surface field mill rule (henceforth referred to as the field mill rule), states that if the electric differential inferred using the field measurement between the sky and ground equals or exceeds the absolute value of 1500 Volts per meter (V m$^{-1}$), then a launch is not authorized unless there are no clouds in the sky (NASA 2017). However, when the field mill measures between 1000-1499 V m$^{-1}$ (in absolute value), different caveats exist. These caveats, however, center on the transparency and the temperature therein of the surrounding clouds. The surface electric field mills can also be used with other LLCC rules to allow launch that would otherwise not be allowed due to other potentially threatening weather phenomena.

The historical context of the field mill rule began with the AC 67 destruction. In the aftermath of this catastrophe, Congress, the United States Air Force, and NASA created independent review panels to determine the reason for the mishap. NASA's independent panel dubbed the 'Busse Panel' centered its review on the CCAFS/KSC's use of the electric field mills (Merceret et al. 2010). They found that after the launch, the weather officer observed that the field mill reading indicated ~ 3000 V m$^{-1}$, signifying a high potential for triggered lightning. This panel also observed that the day after the AC 67 accident, the National Weather Service, which warned against a launch the day before, observed the electric field mill data on the launch day and determined that the readings indicated a dangerous atmospheric electrical environment. Overall, the 'Busse Panel' recommended to officially institute electric field mill measurements as a means of examining lightning potential. Because of these panels, LAP officially created the Surface Electric Fields rule (Merceret et al. 2010).

The CCAFS/KSC has 31 field mills encompassing launch facilities that monitor the surrounding electrical charge and play a pivotal role in monitoring conditions during a space vehicle launch. Figure 1 displays the relative placement of these electric field mills to measure electricity in the atmosphere. The main component of the field mill, the cylindrical object seen at the top in Figure 2, contains a motor that spins a rotor blade at 2,500 revolutions per minute. The field mill records measurements of the ground's electrical activity and records them at a frequency of 50 hertz (3000 readings per minute). It uses these measurements to infer what the electrical potential is in the clouds since the ground charge remains opposite of the charges in the clouds. Therefore, the field mill records these measurements in volts per meter because the field mill stands one meter off the ground and measures the ground's charge there. The field mill sends these measurements as one-minute means, via land line, to the weather team at the CCAFS/KSC.

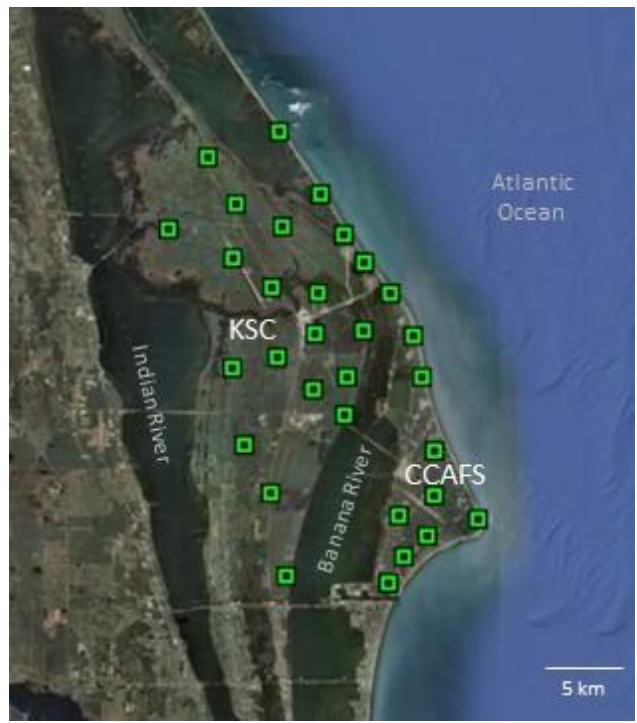

FIG. 1. The surface field mill placement about the CCAFS/KSC facility in Florida.

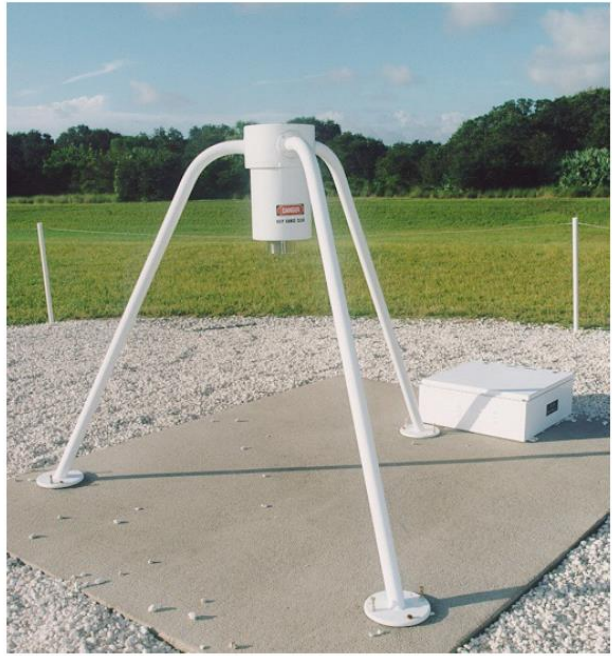

FIG. 2. One of the surface field mill machines located at the CCAFS/KSC. Photo courtesy of NASA (2006).

Studies have demonstrated varying levels of correlation between field mill readings and the likelihood of a lightning strike (Beasley et al. 2008; Murphy et. al. 2008; Montanyà et. al. 2008; Aranguren et. al. 2009; da Silva Ferro et. al. 2011). Additional studies have shown various correlations between field mill readings and the likely presence of thunderstorms and electrified clouds (Liu et al. 2010; Mach et al. 2010, 2011; Blakeslee et al. 2014). Long-term prediction of lightning can be problematic, in comparison to short-term, due to a complexity of meteorological variables to include real-time monitoring of environmental conditions (Hill 2018; Wiston and Mphale 2018; Speranza 2019; Cheng 2020). Due to space operations ramping up considerably for the CCAFS/KSC (Gold et al. 2020), a useful tool for long-range planning would be to provide the 45 WS with a climatology of LLCC violation. Unfortunately, an LLCC climatology has been difficult to create, primarily due to the relatively short periods the LLCC are evaluated and the widely varying dates and times of the launches (Goetz 2000; Muller 2010; Strong 2012; Chafin et al. 2020). The LLCC are sufficiently complex that their violation status is only available during launch windows. As a surrogate, we present a climatology of the field mill rule in the LLCC by using historical data over 18 years. These patterns can then be used as a long-range forecasting tool to establish a baseline estimate of either natural or triggered lightning during the launch window of a space vehicle for any particular hour for a given day and month. These estimates can then be further modified with other forecasting events more amenable to shorter prediction windows.

## 2. Data and Methodology

The creation of the requisite database to build a climatology of the field mill rule as described within the LLCC entailed a two-step process. First, we required the raw data from the 31 field mills surrounding CCAFS/KSC. Secondly, we needed to convert the raw values to a Green, Amber, Red (GAR) designation depending on whether the field mill registered a 1-minute mean absolute value of less than 1000 m$^{-1}$, from 1000 m$^{-1}$ to 1499 m$^{-1}$, or 1500 m$^{-1}$ and greater. This is in keeping with the various thresholds of the surface field mill rule within the LLCC.

With respect to the first step, raw field mill values were obtained from the Global Hydrology Resource Center (GHRC) and the KSC weather archives. The GHRC contains an archive of the CCAFS/KSC's field mill readings from August 1, 1997, to October 13, 2012, while the data from the KSC archive contains field mill readings ranging from January 24, 2013, to the present. However, Lucas et al. (2017) had previously pulled data from both sources and combined them for their own study. In total, their database spans from August 1, 1997, to August 2, 2015, and provides 1-minute means for the 50 hertz field mill readings. We obtained a copy of their raw dataset for our study.

Upon inspection though, the dataset contained several weeks and months of missing data. For continuity from year to year, we excluded February 29th for consideration in our analysis. To address the largest contiguous missing portion, October 13, 2012, through January 24, 2013, we contacted the 45 WS. They were unable to provide the missing 1-minute means in question; however, they were only able to provide 5-minute mean data. Consequently, we used the same 5-minute mean value for each block of 5 minutes of missing data for which the mean corresponded. These three data sources (the two merged by Lucas et al. (2017) and our data retrieval from 45 WS) provide the composition of the initial raw data for our study. All the raw data is in Greenwich Mean Time (GMT) and converted to absolute value in keeping with the surface field mill rule of the LLCC. For all the data processing and subsequent analyses documented in this paper, we used Microsoft Excel 2019 and the JMP Pro 15 statistical software package.

Before converting the data to the GAR designations, we checked for numerical anomalies within the raw data and any remaining data gaps. Specifically, 1997 contained field mill voltage readings for only 3 days in August, 8 days in November, and 23 days in December. Due to this extreme data sparsity and the fact we were unable to acquire the missing data from the 45 WS, we eliminated 1997 from consideration and commenced with 1998.

Another anomaly included recorded occurrences of absolute field mill readings at or above 12,000 V m$^{-1}$, and at times even above 19,000 V m$^{-1}$, which remained constant for several minutes. After consultation with the 45 WS, we removed questionable anomalies that appeared to have the effect of being suddenly spiked (with respect to voltage measurement) and remaining "spiked" for sometimes an hour or longer. This data processing step eliminated a minor 0.004% of all 1-minute field mill readings.

The resultant processed raw dataset recorded the majority of every field mill voltage reading for every minute of every day from January 1998 through August 2015, approximately 312 million rows of data. However, a 1-minute mean represented an unrealistic time window for an operational meteorologist to monitor in practice. Therefore, the CCAFS/KSC indicated that they did not require a field mill climatology based upon 1-minute increments; instead, they preferred 1-hour increments. Consequently, we calculated the hourly mean of the 60-minute means as well as the maximum value of the 60-minute means since we were uncertain whether the hourly mean or maximum field mill value would be more appropriate in establishing a climatology of the field mill rule within the LLCC.

The next step in preparing the data entailed ascertaining whether we needed to account for the values of individual field mills or if the entire system could be reasonably modeled by either the mean of all the field mills or the maximum value of all the fields mills (for each hour). Additionally, we needed to determine if the specific year influenced field mill readings. That is, we needed to assess if voltage readings have been increasing or decreasing over time, after accounting for the monthly and hourly effects.

Using voltage readings of the field mills as the dependent variable, we conducted an Analysis of Variance (ANOVA) with the independent variables of year, month, and time (hourly diurnal effect) based on consultation with the 45 WS. Additionally, we considered a temporal effect of voltage readings by also incorporating the previous hour's voltage reading. It is important to note that the goal of this step was not to build a predictive regression model of voltage readings but to determine the significance of the year effect on voltage readings relative to the other variables in the analysis. Whether we incorporated the previous hour's field mill voltage readings or not, year had the smallest relative effect on field mill voltage readings with the next variable being at least 12 times more associated with voltage readings. Such results are in keeping with Lucas et al. (2017) who also noted the weak effect of year on field mill voltage readings.

Besides investigating an annual effect on voltage readings, we also considered if specific groupings of field mills reported similar voltage readings. If they did, then we could reduce the number of field mill values we had to account for within field mill climatology. Lucas et al. (2017) had suggested that readings from inland field mills were more similar to one another in comparison to the outer coastal field mills. This is expected since fracturing droplets from breaking waves can generate electric charge which is then advected by local wind over the coastal field mills. Thus, we performed Principal Component Analysis (PCA) on the correlations of the voltage readings among the 31 field mills to determine clusters. A PCA seeks to determine which linear combination of variables (in our case individual field mill readings) best explains the variability of the entire dataset.

The PCA did not indicate a difference between coastal field mills and inland field mills as suggested by Lucas et al. (2017). Furthermore, the PCA indicated that the first principal component of the entire field mill system accounted for approximately 79% of the entire system's variability. This percentage was the same whether we analyzed hourly means or hourly maxes. The remaining principal components fell precipitously to such an extent that the next highest principal component explained only 7% of the system's variability. In keeping with the Pareto Principle (also known as the 80/20 rule or principle of factor sparsity (Box and Meyer 1986)), we focused on the single principal component that explained almost 80% of the entire system variability. To determine the appropriate weighting corresponding to the first principal component, we viewed the associated eigenvector. The eigenvector values resembled a relative uniform distribution with rounded values ranging from 0.15-0.19. Such a distribution, given the millions of rows, suggested that either a global mean or max of the entire 31 field mill system was sufficient to model the field mill climatology of the entire network.

The second step of our methodology required converting the numerical field mill values into a GAR (Green/Amber/Red) classification system. This required adhering to the thresholds established in the current LLCC for the field mill rule. This rule states that, barring no other violations, when a field mill absolute voltage reading is less than 1000 V $m^{-1}$, a launch may occur. However, after 1000 V $m^{-1}$, more restrictions apply that require further checking before authorizing a launch. At 1500 V $m^{-1}$, the LLCC does not permit launch unless there are no clouds in the sky. These delineations result in the following classification for our study (values are in absolute value): Green—less than 1000 V m$^{-1}$; Amber—greater than or equal to 1000 V m$^{-1}$ but less than 1500 V m$^{-1}$; Red—greater than or equal to 1500 V m$^{-1}$.

With respect to the actual conversion from numerical to categorical data, we used a dichotomous (0/1) coding scheme. For example, if for a particular hour, the absolute value from a field mill registered less than 1000 m$^{-1}$, then we coded that hour as a '1' for Green and a '0' for both Amber and Red. Conversely, if the value had been 1700 m$^{-1}$, then that hour was a '0' for both Green and Amber but a '1' for Red. Since the prior PCA indicated using either the hourly max or the hourly mean of the entire field mill system, our dichotomous conversion operated on the entire system and not just one particular field mill. Therefore, a particular hourly mean coded as a '1' for Amber implies that for that hour the mean of the entire field mill system was equal to or greater than 1000 m$^{-1}$ but less than 1500 m$^{-1}$. Similarly, a particular hourly max coded as '1' for Red implies that the maximum voltage reading for the entire field mill system for that hour was equal to or greater than 1500 m$^{-1}$.

Having converted the voltage readings into a system of zeros and ones corresponding to whether an hour of the entire field mill system was designated Green, Amber, or Red, we now determined the likelihood of a particular hour being considered low risk, moderate risk, or high risk for violating the field mill rule within the LLCC. Since the previously discussed ANOVA indicated that year had a weak effect on the voltage readings of the field mills, we averaged the zeros and ones for a particular hour to derive the percentage for Green, Amber, and Red, respectively. Since potential differences may have existed in system performance using system means and maxes, primarily influenced by field mill outliers, we calculated these percentages separately. These calculated percentages therein established the climatology of the field mill rule within the LLCC and provided a long-term perspective which the 45 WS can further refine as the launch window approaches. The next section reveals the patterns noted by the climatology and makes a recommendation between the system maxes or means for forecast consideration.

3. Results and Conclusions

Similar to the ANOVA utilized to ascertain the relative effect of year on field mill voltage readings, we conducted an ANOVA to determine whether a climatology based upon hourly means or hourly maxes better associated with the GAR percentages. We used the coefficient of determination, $R^2$, to compare the two (given the very large sample sizes involved) analyses because the largest absolute difference between the reported $R^2$ and the adjusted $R^2$ was 0.0004. The response variable for either the hourly means or maxes was the respective GAR percentage; therefore, we needed to conduct three ANOVAs (Green/Amber/Red). The explanatory variables consisted of the GAR percentages for the previous two hours as well as the hour and month. Table 1 displays the resultant $R^2$ values for both the hourly means and maxes. For both the Green and Red percentages, the hourly max climatology provided a better predictor than the hourly means, with values of 0.711 and 0.738, respectively. The exception came when looking at the Amber percentages. In this instance, the hourly means provided a better climatology $R^2$ value. This result needs to be tempered by the fact that this value was only 0.173.

TABLE 1. Resultant $R^2$ values for the hourly system max or means when investigating the Green, Amber, and Red (GAR) criteria for the field mill rule within the LLCC. The ANOVA response consisted of the respective GAR percentage, while the explanatory variables consisted of the two previous hours GAR percentages, month, and hour effect. Numbers rounded to two significant digits.

| Max/Mean | Red | Amber | Green |
|---|---|---|---|
| Max | 0.74 | 0.12 | 0.71 |
| Mean | 0.42 | 0.17 | 0.56 |

One of the plausible reasons for the Amber value reflecting a low $R^2$ is due to the low sample size of the Amber criteria flagging in general. Within the raw dataset, the Amber criteria existed primarily when atmospheric electricity transitioned from low to moderate, which was usually during sunrise. This so called 'sunrise' effect was also noted by Lucas at el. (2017). Consequently, in the original 1-hour block max and mean categories, the Amber criteria accounted for only approximately 0.8% of total values. In comparison, the Red criteria accounted for 3.7% and 1.5% of all total values for these max and mean databases, respectively.

In addition to Table 1 suggesting that the hourly maxes provided a better climatology for the field mill rule, the requirement remains that the sum of all GAR percentages for a particular hour must sum to 100%. That is, even though the hourly means seem to explain more of the system variability than hourly maxes for Amber criteria, it is not possible to add likely mean event percentage to corresponding max percentages for the Red or Green criteria because these three percentages would not add to 100%. Consequently, for the remainder of the analysis, we used the max hourly percentages for all three criteria.

Figures 3 through 7 highlight the monthly GAR percentages and depict the general climatology pattern of the field mill rule within the LLCC. All presented times are in GMT. In all graphs, approximately around 1300 GMT (~ 0900 EDT), one can see the increase of the Amber percentage, which reflects the 'sunrise effect' noted by Marshall et al. (1999) and Lucas et al. (2017). Additionally, one can note that there appears to be a two-season pattern during the year with respect to the Green and Red percentages. As noted by Gold et al. (2020), the warm season appears to span the months from June to September, while the cold season spans the remaining months.

During the warm season, the Red percentages indicate an increased likelihood of having a higher field mill reading and thus higher chance of violating the field mill rule within the LLCC. Conversely, the decreased Green percentages indicate a lowering percentage of not violating the field mill rule. On average, 2000 GMT (~ 1600 EDT) is when the highest likelihood of violating the field mill rule starts to occur. Not surprisingly, this coincides with the timing of summer thunderstorms around the CCAFS/KSC and in Florida in general (Matsui et al. 2010). We split the months of June and September into two, 2-week periods in order to refine the transition between cold and warm seasons (June) and the reverse (September). However, based on the resultant graphs, there appeared no mid-month transition.

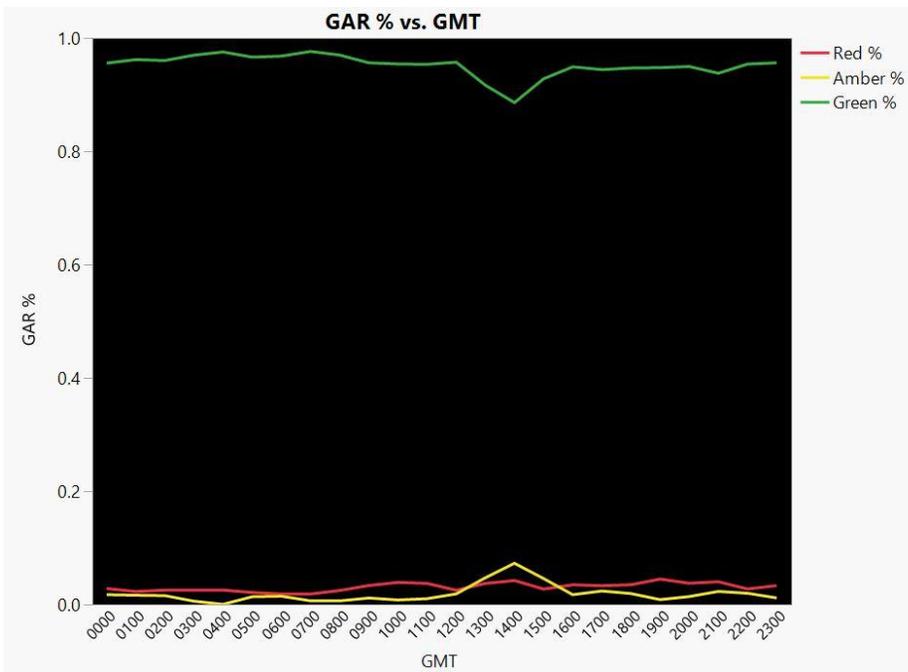

FIG. 3. Typical presentation of system-wide climatology pattern for the likelihood of violating the field mill rule of the LLCC for months November through February by hour. Times are in GMT.

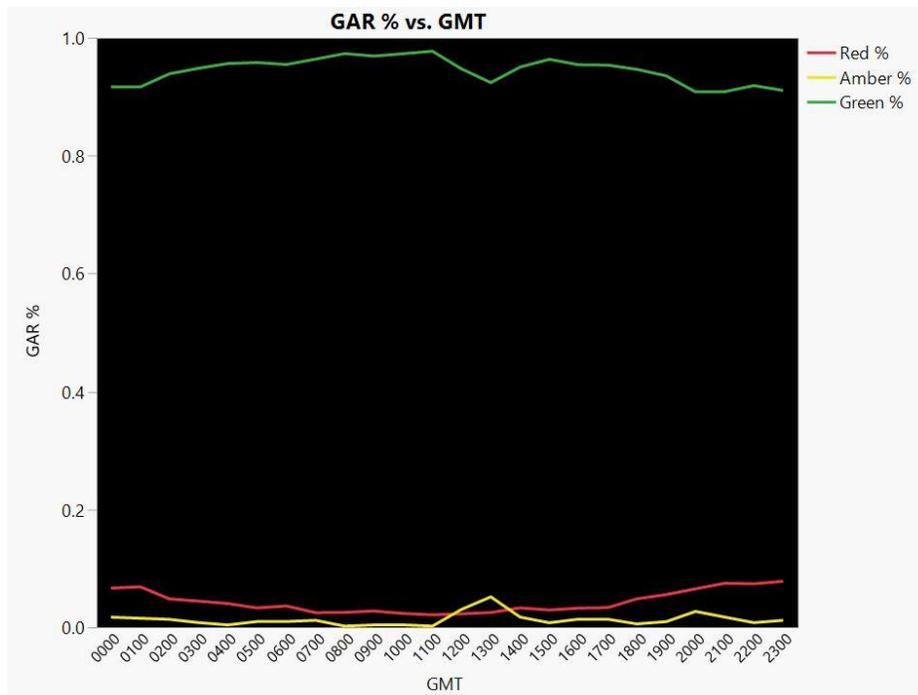

FIG. 4. Typical presentation of system-wide climatology for the likelihood of violating the field mill rule of the LLCC for March and April by hour. Times are in GMT.

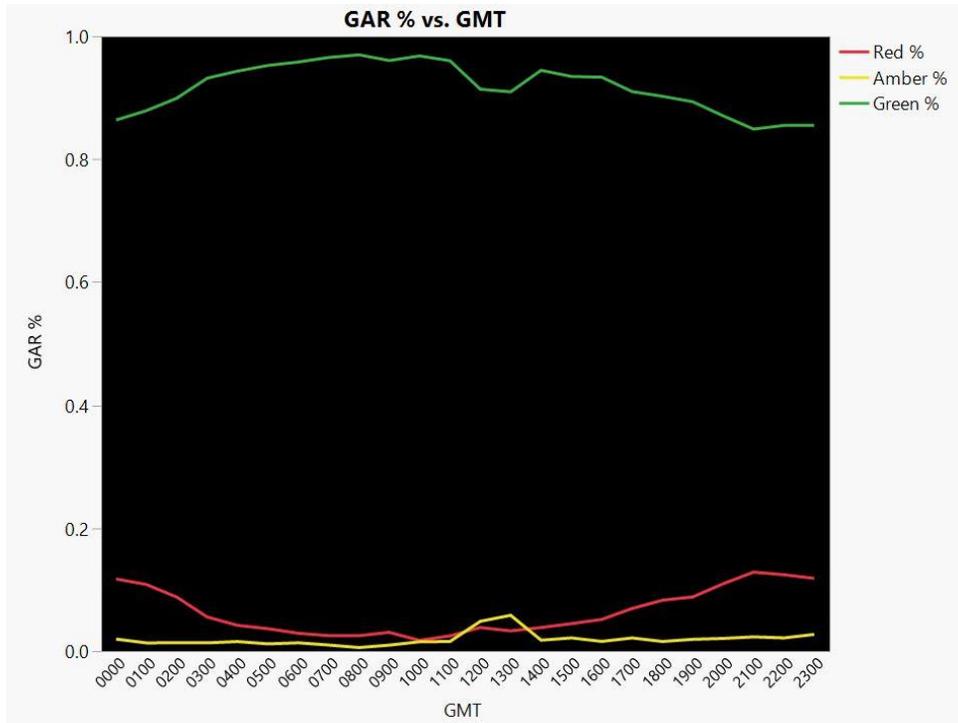

FIG. 5. Typical presentation of system-wide climatology for the likelihood of violating the field mill rule of the LLCC for May and October by hour. Times are in GMT.

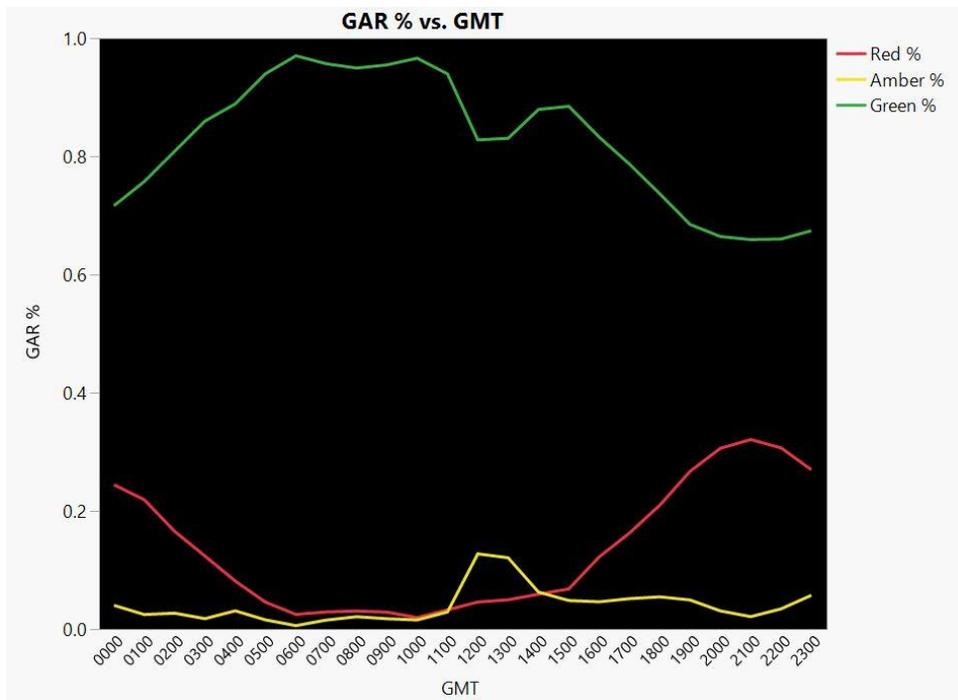

FIG. 6. Typical presentation of system-wide climatology for the likelihood of violating the field mill rule of the LLCC for June, July, and August by hour. Times are in GMT.

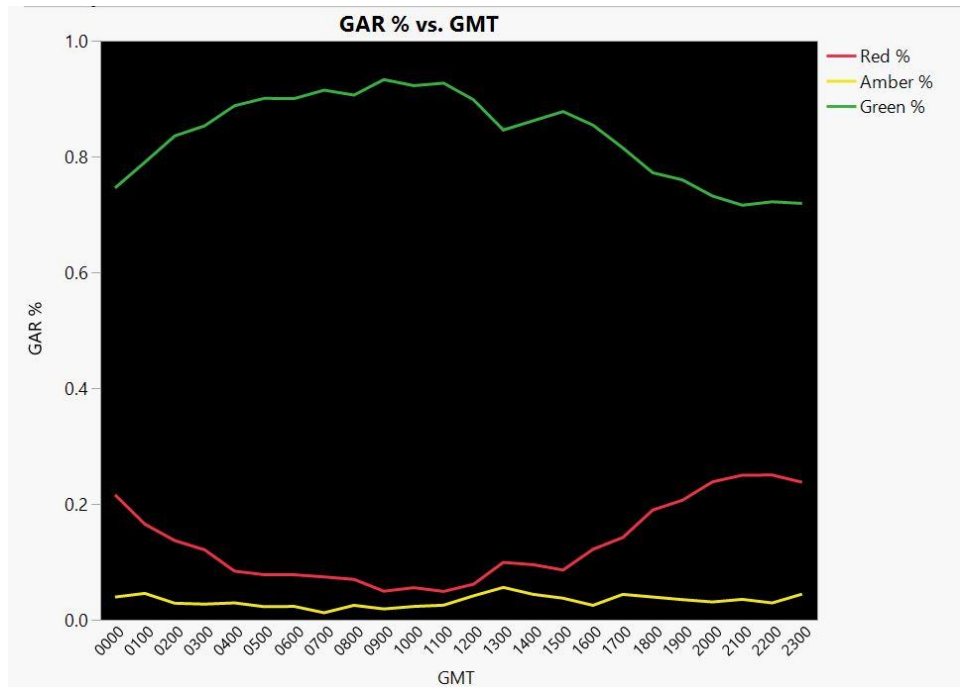

FIG. 7. Typical presentation of system-wide climatology for the likelihood of violating the field mill rule of the LLCC for September by hour. Times are in GMT.

One method to validate the field mill climatology presented in this article involves comparing the modeled patterns to similar assessments in other locales. Figures 8-10 from Krider et al. (2006) show the likelihood of experiencing lightning in the three different regions. Since field mills are designed to detect electrical fields, one would expect the climatology described in this paper to somewhat reflect the likelihood of seasonal lightning patterned elsewhere. When comparing Figures 3-7, which show the climatology of experiencing a low field mill voltage reading (the Green percentage) to Figures 8-10, the graphs are remarkably similar. In all cases, one can see that the proposed climatology reflects a lower Green percentage trend during the warm season months and a much higher one in the cold season months. This comparison suggests the climatology patterns ascertained from the field mills at Cape Canaveral are comparable to areas quite geographical dissimilar with seasonal lightning.

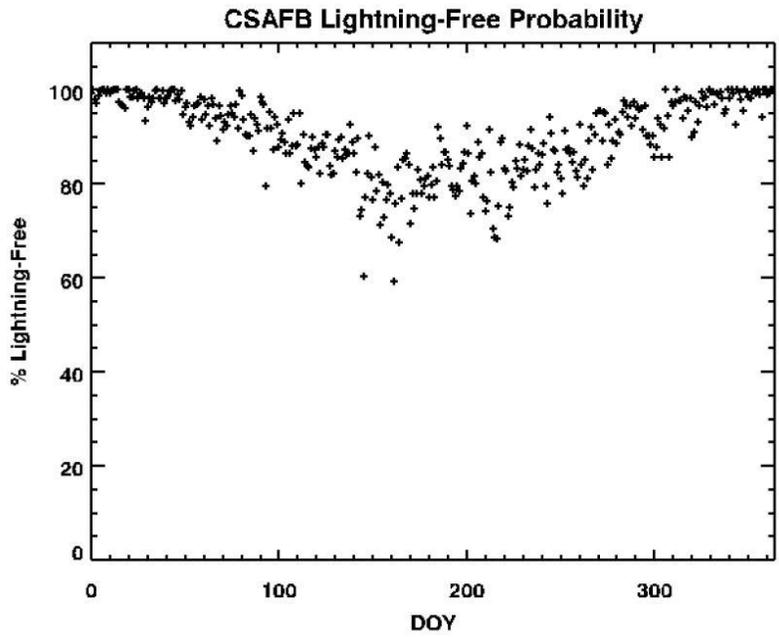

Figure 3-33. Probability that a DOY will be lightning free at Clinton-Sherman AFB.

FIG. 8. First sample distribution in Krider et al. (2006) for comparing to Figures 3-7. DOY = Day of the Year.

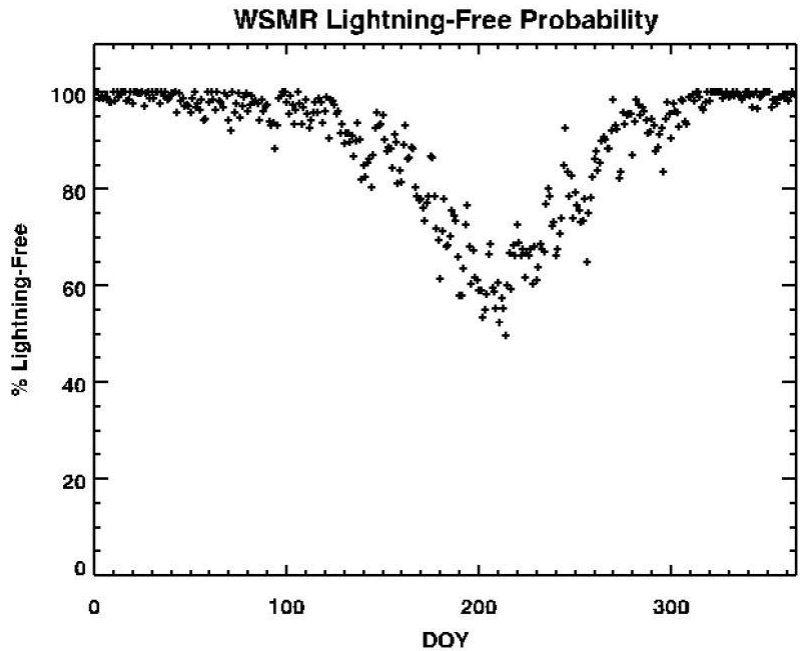

Figure 3-31. Probability that a DOY will be lightning free at White Sands Missile Range.

FIG. 9. Second sample distribution in Krider et al. (2006) for comparing to Figures 3-7. DOY = Day of the Year.

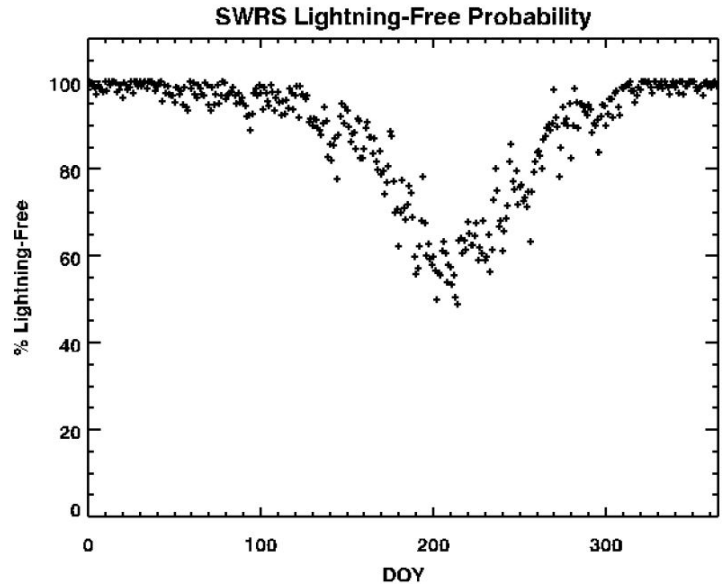

Figure 3-32. Probability that a DOY will be lightning free at Southwest Regional Spaceport.

FIG. 10. Third sample distribution in Krider et al. (2006) for comparing to Figures 3-7. DOY = Day of the Year.

With respect to assessing how well the developed climatology corresponds to the LLCC rules, we compared the GAR percentages presented by the developed climatology to successful launches at Cape Canaveral. From January 2005 to January 2020, there were 159 successful CCAFS/KSC launches. According to NASA (2017), this means that every LLCC was met, including the field mill rule. Therefore, we know that during these 159 instances, CCAFS/KSC did not witness the possibility of either a natural or triggered lightning strike.

Using these 159 instances and specific launch times, we examined the corresponding violation percentage in the field mill climatology we created. Of the 159 instances, 140 launches (88%) occurred during hours when the percentage of violating the field mill rule remained at or below 15%. Therefore, according to our climatology, most successful launches corresponded with a low chance of violating the field mill rule and hence the LLCC. These results suggest that the overall field mill climatology pattern can provide a long-term planning tool for forecasting a field mill violation with respect to space vehicle launches.

In addition to the above verification, the Launch Weather Officers at 45 WS reviewed the field mill LLCC climatology. These Launch Weather Officers have nearly a century of combined experience evaluating the LLCC at CCAFS/KSC. The field mill LLCC climatology for $\geq 1500$ Vm$^{-1}$ (absolute value) shows the expected day-of-year and diurnal patterns. For example, there are lower violations rates and less diurnal variation during the dry season (Oct-Apr) when most of the LLCC violations are due to relatively infrequent fronts affecting the area. Conversely, there are higher values and very pronounced diurnal variation during the wet season (late May-mid Sep) when most of the LLCC violations are caused by locally driven low-altitude boundary interactions and occur almost daily. Intermediate patterns are seen during the seasonal transition months of May and Sep. In addition to showing the expected day of year and diurnal patterns, the magnitude of the $\geq 1500$ Vm$^{-1}$ (absolute value) are close to the magnitudes of the launch forecasts. This reinforces the result that the field mill LLCC can be useful operationally at CCAFS/KSC.

As the number of launches of space vehicles increase at CCAFS/KSC, the need for long-term planning becomes even more critical. This field mill LLCC climatology can be one tool to help improve that long-term planning. We used 18 years of data from 31 field mills surrounding the CCAFS/KSC to generate a climatology pattern for the likelihood of a field mill rule violation. Our results suggest that the hourly maximum voltage value of the 31 field mills best correlates with the Green and Red percentages (not violating and violating the field mill rule, respectively). For the Amber criteria, the system mean seems to perform better. However, such occurrences generally were associated with the sunrise effect. Lastly, our results showed little to no yearly effect with respect to voltage readings and a two-season field mill voltage pattern around the geographical location of the CCAFS/KSC. Since our climatology patterns mirrors similar results from Krider et al. (2006), we suggest adopting this methodology in other areas where seasonal lightning is a concern.

## 4. Future Work

Initial use of this field mill LLCC climatology by 45 WS has been positive. Future development will include allowing start times every 30 min for any day of the year and the probability of violation for the next 0.5 hr, 1.0 hr, 1.5 hr, 2.0 hr, 2.5 hr, 3.0 hr, 3.5 hr, and 4.0 hr. This will allow finer temporal resolution and cover the duration of most launch windows as compared to the current climatology with 1 hr start times and the probability of violation over the next 1.0 hr. In addition, a 2D histogram of the probability of violation by hour of day and day of year will be created for training of 45 WS Launch Weather Officers and educating launch customers.

## Acknowledgments

Many thanks to friends and colleagues who assisted in this research and in particular the reviewers for their constructive comments.  This work was partially supported by the 45th Weather Squadron.